\documentstyle[12pt,epsf]{article}

\begin{document}
\renewcommand{\thefootnote}{\alph{footnote}}

\begin{titlepage}

\vspace*{1cm}

\begin{center}
\large\bf
MULTIGLUON AMPLITUDES IN THE HIGH-ENERGY LIMIT\footnote{Talk presented at the
${\rm XXX}^{th}$ Rencontres de Moriond on ``QCD and High Energy Hadronic
Interactions", Les Arcs, France, March 19-26, 1995.}
\end{center}

\vspace*{0.5cm}

\begin{center}
Vittorio Del Duca \\
Deutsches Elektronen-Synchrotron \\
DESY, D-22603 Hamburg , GERMANY
\end{center}

\vspace*{1cm}

\begin{center}
\bf Abstract
\end{center}

\noindent

We give a unified description of tree-level multigluon amplitudes in the
high-energy limit. We represent the Parke-Taylor amplitudes and the
Fadin-Kuraev-Lipatov amplitudes in terms of color configurations that are
ordered in rapidity on a two-sided plot. We show that
for the helicity configurations they have in common the
Parke-Taylor amplitudes and the Fadin-Kuraev-Lipatov amplitudes coincide.

\end{titlepage}

\baselineskip=0.8cm

Amplitudes with multi-parton final states have several applications
in high-energy hadronic interactions. They are used in computing:
\begin{itemize}
\item[] multi-jet events, which are of phenomelogical interest because they
appear as background to top-quark and electroweak-boson production and to
eventual Higgs-boson production and new-physics signals;
\item[] scaling violations in DIS at small $x_{bj}$ \cite{haut};
\item[] azimuthal-angle correlations in dijet production at large rapidity
intervals \cite{terry};
\item[] dijet production with large rapidity gaps \cite{phil}.
\end{itemize}
Multi-parton amplitudes are also of interest {\sl per se} because they yield
the radiative
corrections to the total parton cross section, which in the high-energy limit
of perturbative QCD is predicted to have a power-like growth in the parton
center-of-mass energy $\sqrt{\hat s}$ \cite{FKL}.

\begin{figure}[htb]
\vspace*{-4.5cm}
\hspace*{2.5cm}
\epsfxsize=15cm \epsfbox{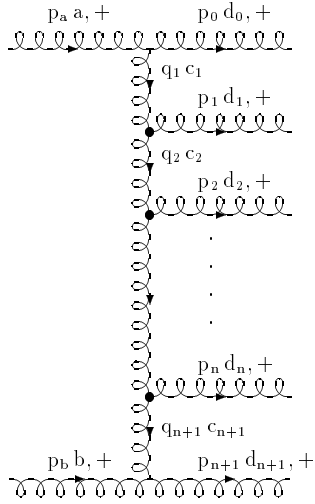}
\vspace*{-10.cm}
\caption{FKL amplitude for fixed gluon helicities. The blobs remind that
Lipatov vertices are used for the gluon emissions along the ladder.}
\label{fig:fkl}
\end{figure}

Multi-parton amplitudes have been computed in the high-energy limit by Fadin,
Kuraev and Lipatov (FKL) \cite{FKL}, who considered the tree-level
production of $n$ gluons in parton-parton scattering in the limit of a
strong rapidity ordering of the produced partons, assuming their
transverse momenta to be all of the same size, $Q$. This kinematic
configuration is termed {\sl multiregge kinematics}. The amplitudes are given
by the exchange of a gluon ladder between the scattering partons
(Fig.\ref{fig:fkl}). FKL made also an ansatz for the leading logarithmic
contribution, in $\ln(\hat s/Q^2)$, of the loop corrections to the
multi-parton amplitudes, to all orders in $\alpha_s$. This changes
the form of the propagators of the gluons exchanged in the $\hat t$ channel,
but preserves the ladder structure of the amplitudes.

On the other hand, {\sl exact} tree-level amplitudes for the production of $n$
gluons have been computed by Parke and Taylor (PT) \cite{pt} in a helicity
basis, for specific helicity configurations of the incoming
and outgoing gluons. In a helicity basis the color structure of the tree-level
amplitudes may be decomposed as a sum over all the noncyclic permutations
of the gluon color flows.
In ref.~\cite{vd} we represented the color flows of the squared PT
amplitudes in terms of color lines in the fundamental representation
of SU($N_c$). Permuting the color flows the color lines appear twisted,
however every configuration may be untwisted
introducing a two-sided plot \cite{bj}. We showed that
restricting the squared PT amplitudes to the multiregge kinematics only
the untwisted configurations with the gluons ordered in rapidity
on the two-sided plot contribute. In ref.~\cite{vd} we worked with
the squared PT amplitudes at leading $N_c$, however due to the incoherence of
the leading $N_c$ term in the color sum of the squared PT amplitudes, the
color flows we consider there are the same as the ones of the PT
amplitudes themselves. Thus, also for the PT amplitudes, for which no
approximation in $N_c$ is made, the leading color configurations are the ones
with the gluons ordered in rapidity on the two-sided plot. Considering
then the sum over the leading color flows of the PT amplitudes in the
multiregge kinematics, we have shown \cite{vitt}
that for the helicity configurations they have in common the PT
amplitudes and the FKL amplitudes are equal. In giving here the outline
of the proof, we follow ref.\cite{vitt}.

We consider the production of $n+2$ gluons of momentum $p_i$, with
$i=0,...,n+1$ and $n\ge 0$, in the scattering between two gluons of momenta
$p_A$ and $p_B$, and we assume that the produced gluons satisfy
the multiregge kinematics, i.e. we require that the gluons
are strongly ordered in rapidity $y$ and have comparable transverse momentum,
\begin{equation}
y_0 \gg y_1 \gg ...\gg y_{n+1};\qquad |p_{i\perp}|\simeq|p_{\perp}|\,
.\label{mreg}
\end{equation}

Tree-level multigluon amplitudes in a helicity basis are known {\sl exactly}
in the {\sl maximally
helicity-violating} configurations $(-,-,+,...,+)$ \cite{pt},
\begin{equation}
M(-,-,+,...,+) = 2^{2+n/2}\, g^{n+2}\,
\sum_{[A,0,...,n+1,B]} {\rm tr}(\lambda^a\lambda^{d_0} \cdots
\lambda^{d_{n+1}} \lambda^b) \, {\langle p_i p_j\rangle^4\over
\langle \tilde{p}_A p_0\rangle \cdots\langle p_{n+1} \tilde{p}_B\rangle
\langle \tilde{p}_B \tilde{p}_A\rangle}\, ,\label{two}
\end{equation}
where the $\pm$ signs on the left hand side label the gluon helicities, the
$\lambda$'s are the color matrices in the
fundamental representation of SU($N_c$) and the sum is over the noncyclic
permutations of the set $[A,0,...,B]$,
$i$ and $j$ are the gluons of negative helicity, and we consider all
the momenta as outgoing. The spinor products $\langle p_i p_j\rangle$ are
defined in ref.\cite{mp}.

In the multiregge kinematics the PT
amplitudes (\ref{two}) for which the numerator is the largest are the ones for
which the pair of negative-helicity gluons is one of the following,
\begin {equation}
(A,B),\quad (A,n+1),\quad (B,0),\quad (0,n+1)\, .\label{neg}
\end{equation}
We focus on the first pair, and fix $p_i = \tilde{p}_A = -p_A$ and
$p_j = \tilde{p}_B = -p_B$ in eq.(\ref{two}). The modifications needed
for the other helicity configurations of eq.(\ref{neg}) may be found in
ref.\cite{vitt}.

\begin{figure}[htb]
\vspace{12pt}
\vskip 0cm
\epsfysize=6cm
\centerline{\epsffile{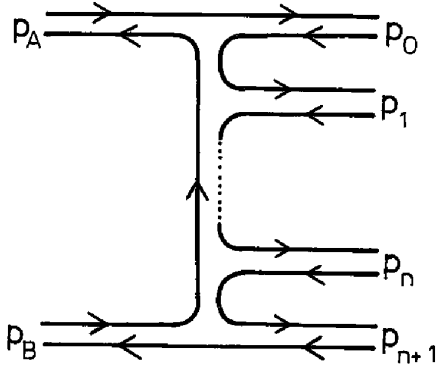}}
\vspace{18pt}
\vskip-1cm
\caption{PT amplitude with color ordering [$A,0,...,n+1,B$].}
\label{fig:one}
\vspace{12pt}
\end{figure}

Then we examine all the color orderings, starting with the ordering
[$A,0,...,n+1,B$]
(Fig.\ref{fig:one}). Computing the string of spinor products in the
denominator of eq.(\ref{two}) we find \cite{vitt},
\begin {equation}
\langle p_A p_0\rangle \cdots\langle p_{n+1} p_B\rangle \langle p_B p_A\rangle
\simeq (-1)^{n+1} \langle p_A p_B\rangle^2 \prod_{i=0}^{n+1} p_{i\perp}\,
.\label{tre}
\end{equation}
Every other color
configuration, for which we keep fixed the position of gluons $A$ and $B$
in the color ordering and permute the outgoing gluons,
gives a subleading contribution, of ${\cal O}(e^{-|y_i-y_j|})$, to
eq.(\ref{two}). We note that untwisting the color lines on a configuration
with permuted outgoing gluons, the color ordering we obtain is different from
the rapidity ordering. Thus the leading color configuration in multiregge
kinematics is the one whose
untwisted lines respect the rapidity ordering (Fig.\ref{fig:one}).

\begin{figure}[htb]
\vspace{12pt}
\vskip 0cm
\epsfysize=6cm
\centerline{\epsffile{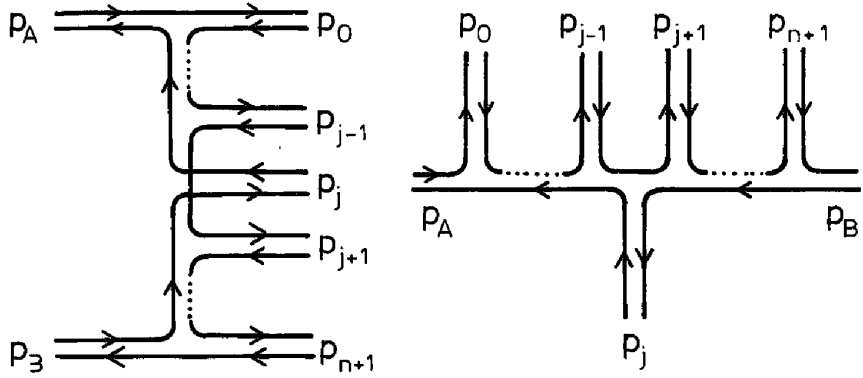}}
\vspace{18pt}
\vskip-1.5cm
\caption{$a)$ PT amplitude with color ordering [$A,0,...,j-1,j+1,...,n+1,B,
j$], and $b)$ its untwisted version on the two-sided plot.}
\label{fig:two}
\vspace{12pt}
\end{figure}

Next, we move gluon $B$ one step to the left and consider the color orderings
[$A,0,...,j-1,j+1,...,n+1,B,j$], with
$j=0,...,n+1$ (Fig.\ref{fig:two}a). Untwisting the color lines, we get
gluon $j$ on the back of the plot (Fig.\ref{fig:two}b). We
compute then the string of spinor products in eq.(\ref{two}),
and we note that the result is independent of which gluon we have taken to the
back of the plot in Fig.\ref{fig:two}b.
As before every permutation of the gluons on the front of
the plot of Fig.\ref{fig:two}b gives a subleading contribution, of ${\cal O}
(e^{-|y_i-y_j|})$, to eq.(\ref{two}). Thus the leading color configurations
are the ($n+2$) configurations whose untwisted lines respect the
rapidity ordering on the front of the plot of Fig.\ref{fig:two}b.

We can then proceed further by moving gluon $B$ one more step to the left.
Taking gluon $B$ all the way to the left, we will have
exhausted all the $(n+3)!$ noncyclic permutations of the color ordering
$[A,0,...,B]$.
Substituting then the leading contributions of the different color
configurations into eq.(\ref{two}), we obtain
\begin{eqnarray}
& & M(-p_A,-; p_0,+;...; p_{n+1},+; -p_B,-) \simeq \nonumber\\ & &
(-1)^{n+1}\, 2^{2+n/2}\, g^{n+2}\, {\hat s}\, {1\over \prod_{i=0}^{n+1}
p_{i\perp}}\, {\rm tr}\left(\lambda^a\,\left[\lambda^{d_0},\left[
\lambda^{d_1},...,\left[\lambda^{d_{n+1}}, \lambda^b\right]\right]\right]
\right)\, ,\label{mrpt}
\end{eqnarray}
where the color orderings which contribute to eq.(\ref{mrpt}) in the
multiregge kinematics are given by the $2^{n+2}$ configurations which respect
the rapidity ordering on the two-sided plot.

The tree-level Fadin-Kuraev-Lipatov amplitude \cite{FKL} for the production
of $n+2$ gluons in the multiregge kinematics is given by,
\begin{eqnarray}
M^{ad_0...d_{n+1}b}_{\nu_A\nu_0...\nu_{n+1}\nu_B} &\simeq& 2 {\hat s}
\left(i g\, f^{ad_0c_1}\, \Gamma^{\mu_A\,\mu_0}\right)\,
\epsilon_{\mu_A}^{\nu_A*}(p_A) \epsilon_{\mu_0}^{\nu_0}(p_0)\, {1\over\hat t_1}
\nonumber\\ &\cdot& \left(i g\, f^{c_1d_1c_2}\, C^{\mu_1}(q_1,q_2)\right)\,
\epsilon_{\mu_1}^{\nu_1}(p_1)\, {1\over \hat t_2} \nonumber\\ &\cdot&
\label{ntree}\\ &\cdot&\nonumber\\ &\cdot& \left(i g\, f^{c_nd_nc_{n+1}}\,
C^{\mu_n}(q_n,q_{n+1})\right)\, \epsilon_{\mu_n}^{\nu_n}(p_n)\,
{1\over \hat t_{n+1}} \nonumber\\ &\cdot& \left(i g\, f^{bd_{n+1}c_{n+1}}\,
\Gamma^{\mu_b\,\mu_{n+1}}\right)\, \epsilon_{\mu_B}^{\nu_B*}(p_B)
\epsilon_{\mu_{n+1}}^{\nu_{n+1}}(p_{n+1})\, ,\nonumber
\end{eqnarray}
where the $\nu$'s are the helicities, the $q$'s are the momenta of the gluons
exchanged in the $\hat t$ channel, and $\hat t_i = q_i^2 \simeq
-|q_{i\perp}|^2$.
The $\Gamma$-tensors and the Lipatov vertex \cite{FKL} are gauge invariant.

Helicity conservation at the production vertices for the first and the
last gluon along the ladder in
eq.(\ref{ntree}) yields the four helicity configurations (\ref{neg}).
We choose then the configuration of
Fig.\ref{fig:fkl}, which corresponds to the first of the configurations
(\ref{neg}). There is however no restriction in eq.(\ref{ntree}) on
the helicities of the gluons produced from the Lipatov vertices along the
ladder.

For consistency we use the representation of the gluon polarization used
in the PT amplitudes \cite{mp},
\begin {equation}
\epsilon_{\mu}^{\pm}(p,k) = \pm {\langle p\pm |\gamma_{\mu}| k\pm\rangle\over
\sqrt{2} \langle k\mp | p\pm \rangle}\, ,\label{hpol}
\end{equation}
where $k$ is an arbitrary reference light-like momentum. For the polarization
of gluons $p_A$ and $p_0$ we choose $p_B$ as reference vector, while for
the polarization of gluons $p_B$ and $p_i$ with $i=1,...,n+1$ we choose $p_A$.
The contraction of the Lipatov vertex with the gluon polarization (\ref{hpol})
in eq.(\ref{ntree}) is \cite{lipat}, \cite{vitt},
\begin{equation}
\epsilon^+(p_i, p_A)\cdot C(q_i,q_{i+1}) = \sqrt{2}\, {q_{i\perp}^*
q_{i+1\perp}\over p_{i\perp}}\, .\label{verc}
\end{equation}
The contractions of the helicity-conserving
tensors with the gluon polarizations are \cite{vitt},
\begin{eqnarray}
\Gamma^{\mu_B\mu_{n+1}}\,\epsilon^{+*}_{\mu_B}(p_B, p_A)\,
\epsilon^+_{\mu_{n+1}}(p_{n+1}, p_A) &=& -{p_{n+1\perp}^*\over
p_{n+1\perp}}\, ,\label{contra}\\
\Gamma^{\mu_A\mu_0}\,\epsilon^{+*}_{\mu_A}(p_A, p_B)\, \epsilon^+_{\mu_0}(p_0,
p_B) &=& -1\, .\nonumber
\end{eqnarray}
We rewrite then the product of structure constants in eq.(\ref{ntree})
as the trace of a product of $\lambda$-matrices,
\begin{equation}
f^{ad_0c_1}\, f^{c_1d_1c_2}\,\cdots f^{c_nd_nc_{n+1}} f^{bd_{n+1}c_{n+1}} =
-2\, (-i)^{n+2}\, {\rm tr}\left(\lambda^a\,\left[\lambda^{d_0},
\left[\lambda^{d_1},...,\left[\lambda^{d_{n+1}}, \lambda^b\right]\right]\right]
\right)\, ,\label{fpro}
\end{equation}
which shows that also for the FKL amplitudes the only configurations
which contribute are the $2^{n+2}$ color
configurations which respect the rapidity ordering on the two-sided plot.
Substituting eq.(\ref{verc}), (\ref{contra}) and (\ref{fpro}) into
eq.(\ref{ntree}), the FKL
amplitude in the helicity configuration of Fig.\ref{fig:fkl} is
in agreement with eq.(\ref{mrpt}), thereby proving that the PT amplitudes and
the FKL amplitudes coincide in the high-energy
limit.

Finally, as remarked in the introductive paragraphs, including the leading
logarithmic
contributions, in $\ln(\hat s/\hat t)$, of the loop corrections to
eq.(\ref{ntree}) modifies the propagator of the gluon of momentum $q_i$
exchanged in the $\hat t$ channel only by a function of $\hat{t}_i$ \cite{FKL}.
Thus the FKL amplitude with the
leading-logarithmic loop corrections retains the ladder structure of
eq.(\ref{ntree}) and the color structure of eq.(\ref{fpro}),
and the leading color configurations are still
the ones whose untwisted lines are ordered in rapidity on the
two-sided plot. Even though this is a simple observation from the standpoint
of the FKL amplitudes, it is far from being obvious when we consider the
color decomposition of multigluon amplitudes in a helicity basis, since the
color structure of the tree-level amplitudes (\ref{two}) does not
describe all the possible color configurations of $n$ gluons, more
configurations appearing in the color decomposition at the loop level
\cite{bk}.

\end{document}